\documentclass[aps,prl,twocolumn,showpacs,showkeys,footinbib,amsmath,amssymb,superscriptaddress]{revtex4-2}

\usepackage{amsmath}
\usepackage{amssymb}
\usepackage{graphicx}
\usepackage{bm}
\providecommand{\vb}[1]{\bm{#1}}
\usepackage{color}
\usepackage{relsize}
\usepackage{braket}
\usepackage{bbold}
\usepackage{txfonts}
\usepackage{epstopdf}
\PassOptionsToPackage{colorlinks=true,
	linkcolor=blue,
	citecolor=blue,
	urlcolor=blue}{hyperref}
\usepackage{hyperref}

\begin{document}
	
\title{Surface Functionalization Enables Two-Dimensional Altermagnetism and Giant Tunnel Magnetoresistance}

\author{Zhou Cui}
\affiliation{Ningbo Institute of Digital Twin, Eastern Institute of Technology, Ningbo 315200, China}

\author{Ziye Zhu}
\email{zyzhu@eitech.edu.cn}
\affiliation{Ningbo Institute of Digital Twin, Eastern Institute of Technology, Ningbo 315200, China}

\author{Bowen Hao}
\affiliation{Ningbo Institute of Digital Twin, Eastern Institute of Technology, Ningbo 315200, China}

\author{Xunkai Duan}
\affiliation{Ningbo Institute of Digital Twin, Eastern Institute of Technology, Ningbo 315200, China}

\author{Xuan Zhou}
\affiliation{Ningbo Institute of Digital Twin, Eastern Institute of Technology, Ningbo 315200, China}

\author{Yali Xie}
\affiliation{Zhejiang Province Key Laboratory of Magnetic Materials and Application Technology, Ningbo Institute of Materials Technology and Engineering,
Chinese Academy of Sciences, Ningbo 315201, China}

\author{Huali Yang}
\affiliation{Zhejiang Province Key Laboratory of Magnetic Materials and Application Technology, Ningbo Institute of Materials Technology and Engineering,
Chinese Academy of Sciences, Ningbo 315201, China}

\author{Baisheng Sa}
\email{bssa@fzu.edu.cn}
\affiliation{Materials Genome Institute, College of Materials Science and Engineering, Fuzhou University, Fuzhou 350108, China.}

\author{Runwei Li}
\affiliation{Ningbo Institute of Digital Twin, Eastern Institute of Technology, Ningbo 315200, China}
\affiliation{Zhejiang Province Key Laboratory of Magnetic Materials and Application Technology, Ningbo Institute of Materials Technology and Engineering,
Chinese Academy of Sciences, Ningbo 315201, China}

\author{Tong Zhou}
\email{tzhou@eitech.edu.cn}
\affiliation{Ningbo Institute of Digital Twin, Eastern Institute of Technology, Ningbo 315200, China}	
\date{\today}

\vspace{1em}

\begin{abstract}
	Two-dimensional (2D) altermagnets (AMs) are highly desirable for ultrafast, stray-field-free spintronics because they combine compensated magnetic order and momentum-dependent spin splitting with the scalability, tunability, and interface compatibility of atomically thin materials. However, practical 2D AMs remain scarce. Rather than relying solely on the search for intrinsic 2D AMs, an appealing route is to transform known 2D antiferromagnets (AFMs) into AMs through symmetry engineering. Here, we propose surface functionalization as a symmetry-guided, nonvolatile chemical switch for realizing this AFM-to-AM transformation. By breaking inversion and out-of-plane mirror symmetries while preserving the rotation symmetry connecting opposite-spin sublattices, single-sided functionalization lifts spin degeneracy and induces altermagnetic spin splitting. Using monolayer FeSe as a representative platform, first-principles calculations show that hydrogenation, oxidation, and fluorination convert spin-degenerate antiferromagnetic FeSe into a d-wave AM with pronounced momentum-dependent spin splitting. At the device level, our transport simulations reveal that the functionalized FeSe monolayer magnetic tunnel junctions exhibit giant tunnel magnetoresistance (TMR) up to $1.87\times10^3\%$, originating from momentum-selective spin filtering between parallel and antiparallel Néel-vector configurations. The strong dependence of TMR on functionalization geometry further demonstrates that surface chemistry provides an effective control knob for altermagnetic transport. Our work establishes a symmetry-to-chemistry-to-device strategy for engineering 2D AMs and developing high-performance altermagnetic spintronic devices. 
	
\end{abstract}

\maketitle

\vspace{1em}

Altermagnets (AMs) have recently emerged as a distinct class of collinear magnets beyond conventional ferromagnets and antiferromagnets (AFMs), combining compensated magnetic order in real space with momentum-dependent spin splitting in reciprocal space~\cite{WuPRB2027,hayami2019momentum,Libor2020Crystal,yuan2020giant,mazin2021prediction,yuan2021:prediction,JunweiliuNC2021,Liu2025,Vsmejkal2022:beyond,Smejkal2022a,Yao2024AMReview,song2025altermagnets,Bhowal2025Review,Jungwirth2026}. This unusual combination enables spin-polarized electronic responses without net magnetization, opening opportunities for ultrafast and stray-field-free spintronic transport~\cite{Gurung2024,PhysRevApplied.23.014013,Samanta2025,PhysRevLett.130.036702,Zhou2025Manipulation}, multiferroics~\cite{Duan2025PRL,zhu2025two,zhu2025emergent,Gu2025PRL,sun2025proposing,cao2024designing,guo2026altermagnetic,UrruPRB2025,PengNPJQM2025,Cui2026AD,jql2-b4c4,m33v-xwn3,Yu2025AM},  topology~\cite{Guo2023npj,PhysRevB.110.064426,feng2025type,zhang2025quantized,Chen2026arXiv,huang2026arXiv}, proximity-driven phenomena~\cite{zhu2025altermagnetic} and spin transistor devices~\cite{Liu2026Altermagnetic,Zhu2025AMSFET}. In particular, two-dimensional (2D) AMs are highly desirable because atomically thin materials provide exceptional scalability, interface engineering capability, and compatibility with van der Waals device architectures~\cite{novoselov2016,xu2025chemical}. Practical 2D AMs, however, remain scarce, creating a central bottleneck for both fundamental studies and device applications of altermagnetism.

\begin{figure}[t!]
	\centering
	\vspace{0.2cm}
	\includegraphics*[width=0.45\textwidth]{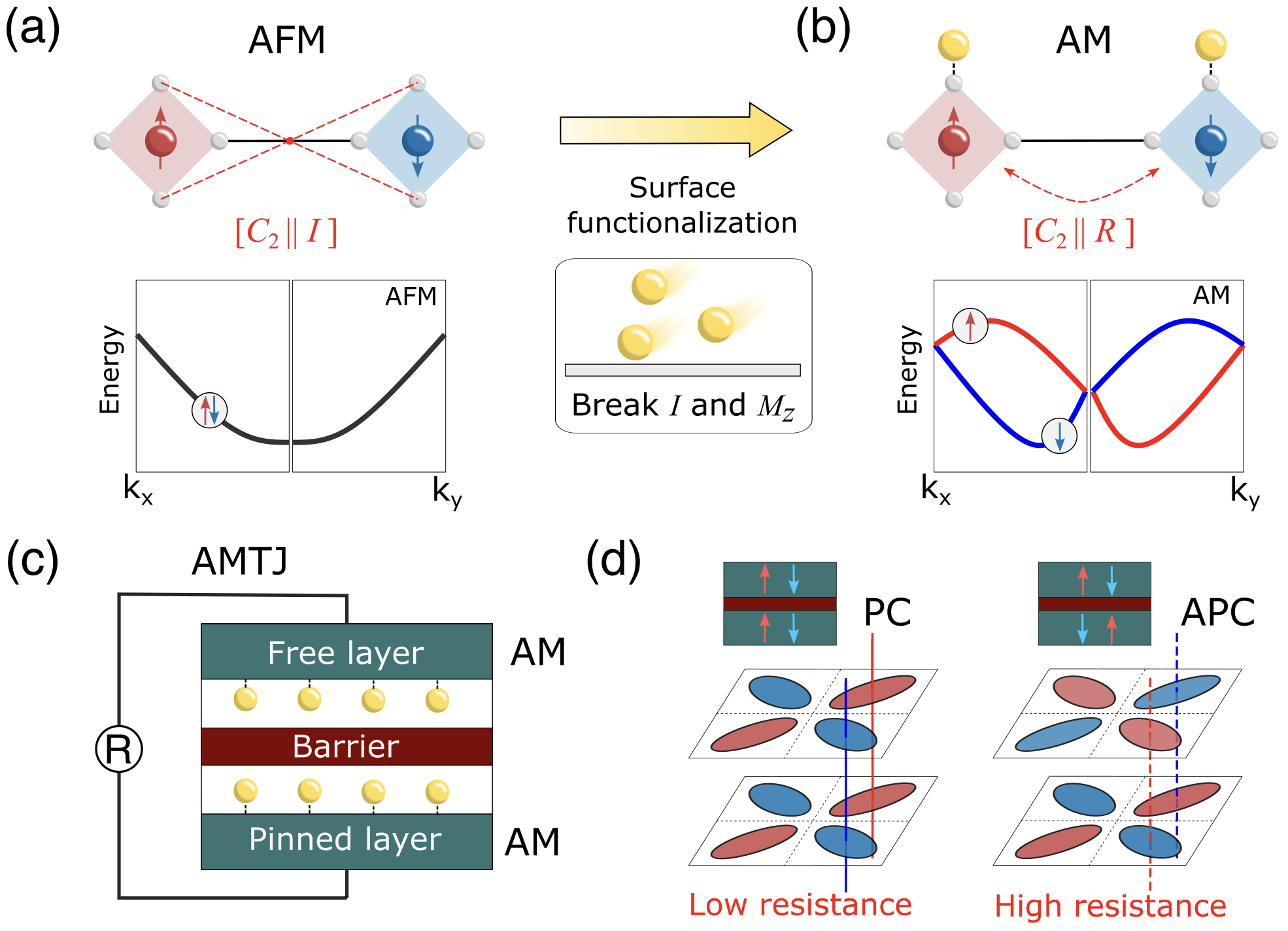}
	\vspace{-0.2cm}
	\caption{Design principle of the surface-functionalization-induced AFM-to-AM transition and altermagnetic tunnel junction (AMTJ).
(a) In the pristine AFM state, opposite-spin sublattices, indicated by red and blue arrows, are related by inversion symmetry $I$, yielding a conventional AFM.
Surface functionalization breaks inversion symmetry $I$ and out-of-plane mirror symmetry $M_z$, producing (b) an AM state in which opposite-spin sublattices are related by rotational symmetry $R$ and exhibit momentum-dependent spin splitting.
(c) Schematic of an AMTJ comprising pinned and free AM layers separated by a spacer.
(d) Momentum-space spin filtering in the AMTJ: parallel N\'eel vectors give matched spin-split Fermi surfaces and a low-resistance state, whereas antiparallel N\'eel vectors cause Fermi-surface mismatch and a high-resistance state.
Red and blue circles denote spin-up and spin-down channels; solid and dashed lines indicate conducting and non-conducting channels, respectively.}
	\label{Figure1}
\end{figure}

\begin{figure*}[t!]
	\centering
	\includegraphics[width=0.95\textwidth]{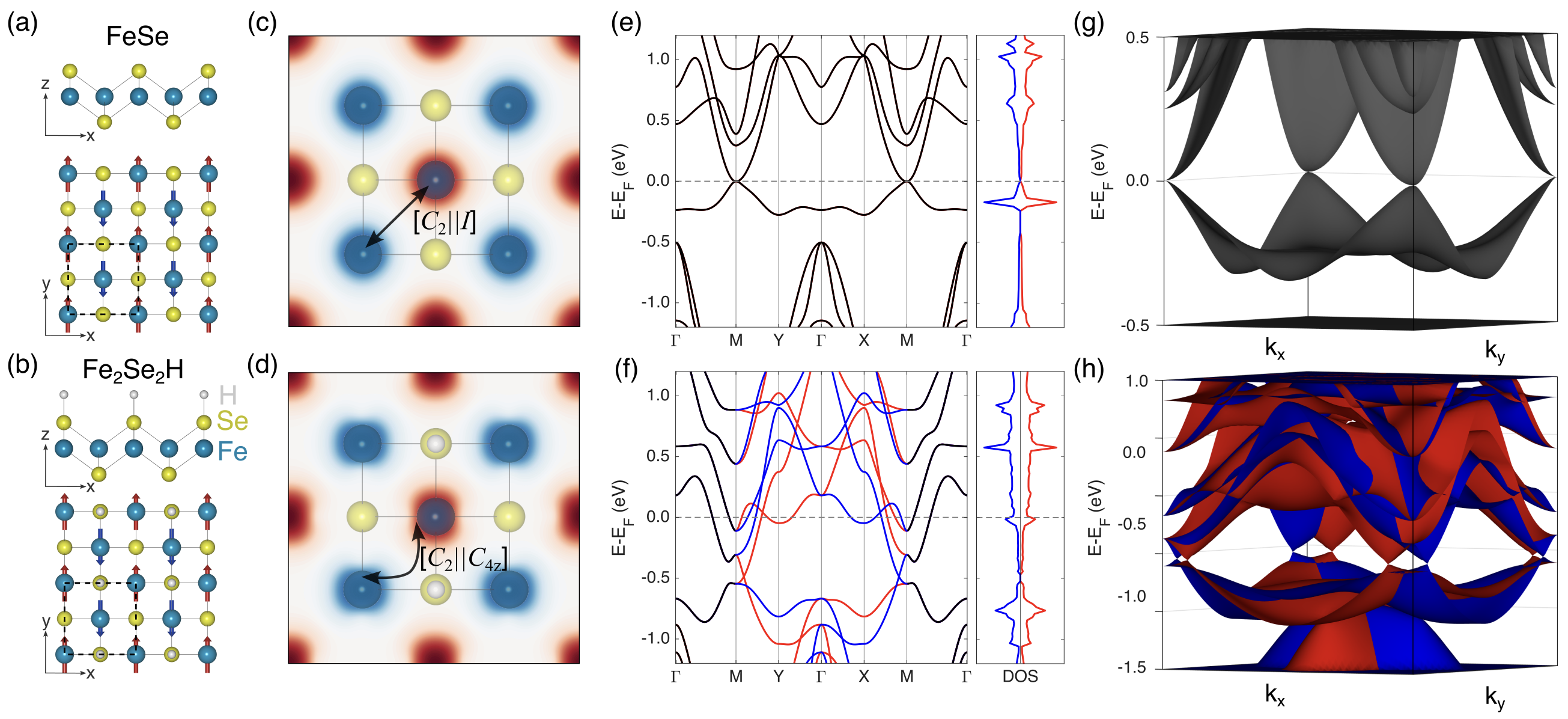}
	\caption{
		Crystal structures of (a) FeSe and (b) the surface functionalization Fe$_2$Se$_2$H monolayers, shown from side and top views. Spin charge densities of (c) FeSe and (d) Fe$_2$Se$_2$H, with red and blue representing spin-up and spin-down components, respectively. Spin-resolved band structures and density of states for (e) FeSe and (f) Fe$_2$Se$_2$H, where black, red, and blue lines denote spin-degenerate, spin-up, and spin-down states. The corresponding spin-resolved density of states is shown on the right, with the Fermi level set to zero. (g) and (h) present the corresponding three-dimensional band structures.
	}
	\label{Figure2}
\end{figure*} 

Beyond the search for intrinsic 2D AMs~\cite{Chen2025Unconventional,k47t-23gp,Gao2025NSR}, a promising strategy is to transform known 2D AFMs into AMs by symmetry engineering. The central requirement is to remove the symmetries that enforce spin degeneracy between opposite-spin sublattices while retaining the crystal-rotation-related symmetry that produces the characteristic momentum-dependent spin splitting~\cite{Vsmejkal2022:beyond,Smejkal2022a,Yao2024AMReview,song2025altermagnets}. Electric fields, strain, bilayer stacking, and twisting have been proposed for this purpose~\cite{mazin2023,PhysRevB.110.174410,liu2024twisted}. Nevertheless, these approaches require volatile external fields, delicate stacking control, or structurally complex heterostructures, raising challenges for stability, reproducibility, and device integration. A nonvolatile, chemically simple, and experimentally accessible route for converting 2D AFMs into AMs is therefore highly desirable.

Here we propose surface functionalization as a symmetry-guided chemical switch for realizing such an AFM-to-AM transformation. In pristine 2D AFMs, opposite-spin sublattices may be connected by inversion symmetry \(I\) or out-of-plane mirror symmetry \(M_z\), yielding spin-degenerate bands and conventional AFM behavior. Single-sided surface functionalization reshapes the local bonding environment and breaks \(I\) and \(M_z\), while the in-plane rotation symmetry connecting opposite-spin sublattices can remain preserved. This symmetry change lifts the spin degeneracy and converts the system into an AM with momentum-dependent spin splitting. Compared with field- or stacking-based approaches, surface functionalization provides a nonvolatile chemical degree of freedom controlled by the functional group, adsorption site, and interfacial geometry~\cite{Du2021,Brill2021}. Established treatments such as hydrogenation~\cite{Barnowsky2024,Albino2024}, oxidation~\cite{Zhang2019}, and halogenation~\cite{Song2023} therefore offer practical routes to chemically engineer altermagnetism in 2D materials~\cite{Garrido2024,Sovizi2023}.

We demonstrate this strategy in monolayer FeSe, a widely studied 2D AFM monolayer with spin degeneracy ensured by \(I\) and \(M_z\) symmetries~\cite{Wei2023FeSePairing,Shi2024,Li2025,Wang2016,Ge2015,luo2022fragile,3kkj-s5jk,mazin2023}. Our first-principles calculations show that the simple hydrogenation and oxidation can break these symmetries and convert FeSe into a $d$-wave AM with pronounced momentum-dependent spin splitting. The device consequence of this chemically induced altermagnetism is particularly striking. We construct Fe$_2$Se$_2$H-based altermagnetic tunnel junctions and evaluate their spin-dependent transport using nonequilibrium Green’s-function simulations. Because the altermagnetic spin polarization is locked to momentum, the tunneling conductance depends sensitively on the relative orientation of the Néel vectors in the two AM electrodes. In the parallel configuration, the momentum-space spin-filtering channels are well matched, producing a high-conductance state. In the antiparallel configuration, the spin-filtering channels become mismatched and the transmission is strongly suppressed, yielding a high-resistance state. This momentum-selective altermagnetic spin filtering produces a giant tunnel magnetoresistance up to \(1.87\times10^3\%\). Moreover, the tunnel magnetoresistance is strongly tunable by the functionalization geometry, demonstrating that surface chemistry provides an efficient control knob for altermagnetic transport. Our work thus establishes a symmetry-to-chemistry-to-device strategy for engineering 2D AMs and suggest broadly applicable route to altermagnetic spintronic devices.

\vspace{0.1cm}
\paragraph*{Design principle.}
From a symmetry perspective, realizing altermagnetism requires that opposite-spin sublattices be related by rotation-related symmetry operations \((R)\), rather than by direct translation \((t)\) or inversion \((I)\)~\cite{Vsmejkal2022:beyond,Smejkal2022a,Yao2024AMReview,song2025altermagnets}. In 2D systems, additional constraints includes the out-of-plane mirror symmetry \((M_z)\) and the twofold rotation about the \(z\) axis \((C_{2z})\)~\cite{PhysRevB.110.174410}. The surface-functionalization-driven AFM-to-AM transition proposed here is also designed based on this symmetry criterion. Here, we consider a specific class of magnetic point groups in which the opposite-spin sublattices are not connected by a pure translation \(t\), as commonly found in buckled lattices or some intrinsically distorted crystals. In such systems, even for a minimal AFM lattice, the opposite-spin sublattices may still be related by inversion \(I\), which can be schematically expressed as \(\{C_2 \parallel I\}\) in Fig.~\ref{Figure1}(a). The band structure therefore remains spin-degenerate, corresponding to a conventional AFM. Surface functionalization lowers the overall symmetry without introducing additional symmetry operations, and generally breaks the \(I\) and \(M_z\) symmetries. Consequently, in the functionalized AFM lattice, only the \(R\)-related symmetry remains to connect the opposite-spin sublattices, denoted as \(\{C_2 \parallel R\}\) in Fig.~\ref{Figure1}(b), thereby lifting the spin degeneracy and giving rise to an AM with momentum-dependent spin splitting.

The proposed symmetry-guided AFM-to-AM transition strategy is particularly suitable for transport devices, as surface functionalization can produce comparatively sizable spin splitting relative to previously proposed strategies. We construct a functionalized altermagnetic magnetic tunnel junction (AMTJ), in which two functionalized AM layers are separated by a nonmagnetic insulating spacer, as shown in Fig.~\ref{Figure1}(c). When the N\'eel vectors of the two AM layers are parallel (PC), their $\bm{k}_\parallel$-dependent spin-polarization patterns are matched, allowing spin-selected electrons to tunnel efficiently through both layers and resulting in low resistance. In contrast, when the N\'eel vectors are antiparallel (APC), the spin-polarization patterns are mismatched, so electrons transmitted through the first AM layer are filtered out by the second one, thereby suppressing the tunneling current and giving rise to a high-resistance state, as illustrated in Fig.~\ref{Figure1}(d). This contrast produces a large tunnel magnetoresistance, defined as \(\mathrm{TMR}=(T_{\mathrm{PC}}-T_{\mathrm{APC}})/T_{\mathrm{APC}}\), where \(T_{\mathrm{PC}}\) and \(T_{\mathrm{APC}}\) are the Brillouin-zone-averaged transmissions for the parallel and antiparallel N\'eel-vector configurations, respectively.

\begin{figure}[t!]
	\centering
	\includegraphics*[width=0.48\textwidth]{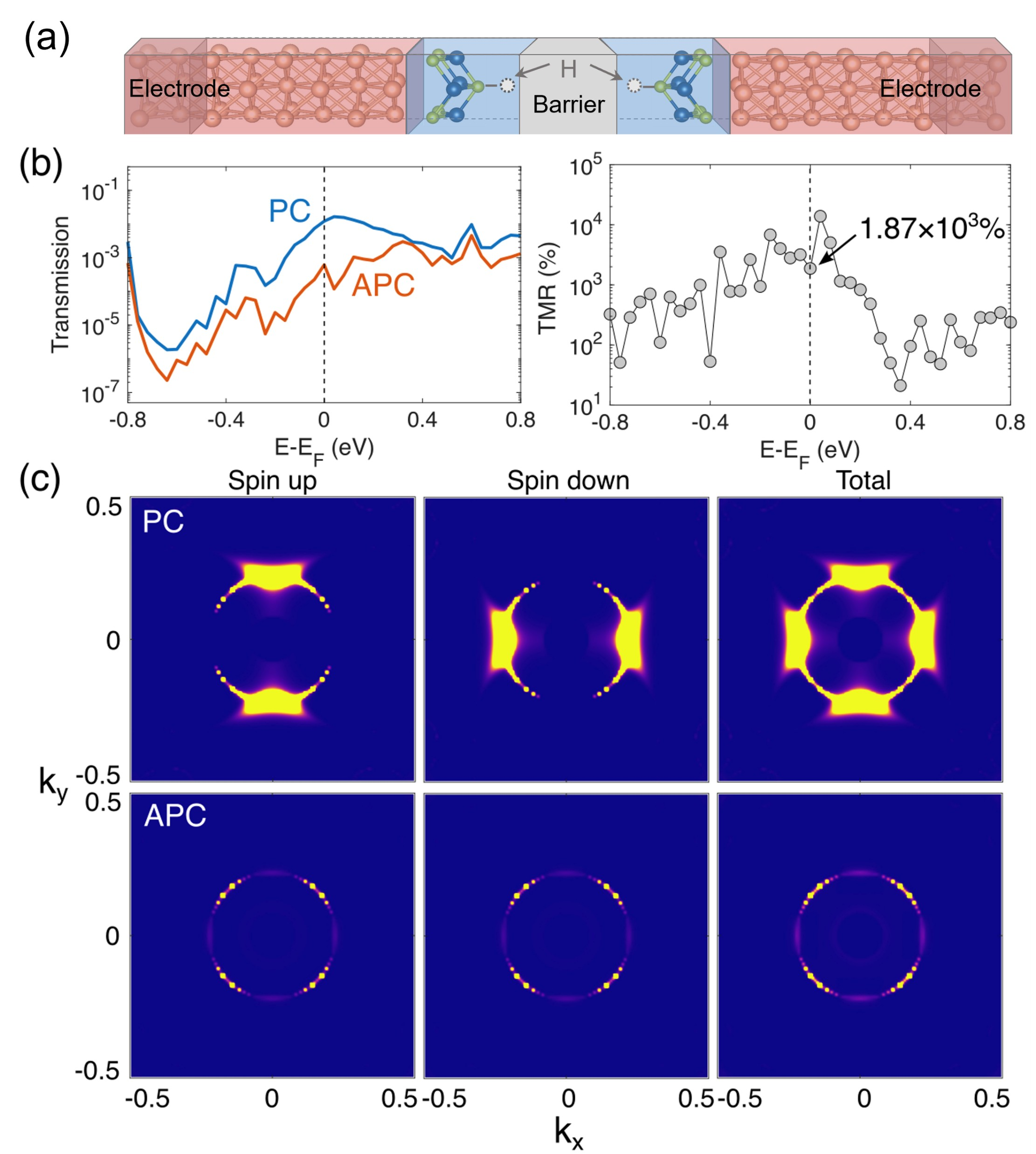}
	\caption{The spin-dependent transport properties of the proposed AMTJ device. (a) Structural models of the Au/Fe$_2$Se$_2$H/GeF$_4$/Fe$_2$Se$_2$H/Au AMTJ. (b) Energy-dependent spin-resolved transmission coefficients for PC and APC configurations, and TMR ratio. (c) $\bm{k}_\parallel$-resolved transmission spectra at the Fermi energy for spin-up, spin-down, and total channels under PC and APC configurations.}
	\label{Figure3}
\end{figure}

\paragraph*{Altermagnetism in functionalized FeSe.}
To verify the symmetry-guided design principle in a realistic material platform, we take iron selenide (FeSe), a prototypical and widely studied material, as a representative case. FeSe is among the most intriguing Fe-based superconductors and has attracted worldwide interest both experimentally and theoretically in recent years~\cite{Wei2023FeSePairing,Shi2024,Li2025,Wang2016,Ge2015}, while theoretical calculations suggest that its monolayer form may host an AFM order~\cite{luo2022fragile,3kkj-s5jk,mazin2023}. Pristine FeSe crystallizes in a tetragonal lattice with \(P4/nmm\) space-group symmetry and is composed of edge-sharing Fe-centered \(\mathrm{FeSe_4}\) tetrahedra [Fig.~\ref{Figure2}(a)]. Its symmetry operations include \(I\), the fourfold rotation \((C_4)\) and \(M\) symmetry, as fully described in Tab.~S1. After considering the AFM order, the magnetic space group becomes \(P4'/n'm'm'\), with reduced symmetry. Nevertheless, the opposite-spin sublattices remain connected by \(I\) or \(M_z\) [Fig.~\ref{Figure2}(c)]. Therefore, pristine FeSe monolayer behaves as a conventional AFM, with spin-degenerate bands throughout the Brillouin zone [Figs.~\ref{Figure2}(e) and (g)].

In contrast, single-sided surface hydrogenation lowers the symmetry of the system. The resulting \(\mathrm{Fe_2Se_2H}\) monolayer has the space group \(P4mm\) and the magnetic space group \(P4'm'm\) [Fig.~\ref{Figure2}(b)]. Surface functionalization selectively breaks the \(I\) and \(M_z\) symmetries while preserving the \(R\)-related symmetries, such as \(C_{4z}\), that connect opposite-spin sublattices [Fig.~\ref{Figure2}(d)]. Such a symmetry configuration satisfies the requirement for altermagnetism. Consequently, the band structure of \(\mathrm{Fe_2Se_2H}\) exhibits momentum-dependent spin splitting [Figs.~\ref{Figure2}{(f) and 2(h)}], and the three-dimensional (3D) band dispersion further reveals its \(d\)-wave character. Similar results in Fig.~S1 are also found for other adsorbed elements, such as \(\mathrm{Fe_2Se_2O}\) and \(\mathrm{Fe_2Se_2F}\), with their dynamic stability confirmed in Fig.~S2. It is worth noting that neither \(\mathrm{FeSe}\) nor \(\mathrm{Fe_2Se_2H}\) monolayers possess the direct \(t\) symmetry. Although they contain the \(C_{2z}\) operation, it does not connect opposite-spin sublattices. Therefore, these systems satisfy the prerequisite of our design principle.

More intriguingly, \(\mathrm{Fe_2Se_2H}\) also exhibits nontrivial topological features. As shown in Figs.~\ref{Figure2}{(f) and (h)}, we identify two pairs of opposite Weyl points near \(-0.5\)~eV, located along the M--Y and X--M paths. When spin-orbit coupling (SOC) is included, these Weyl crossings are gapped and generate pronounced Berry-curvature hotspots with opposite signs in the two spin-polarized sectors, as shown in Fig. S3. Integration of the Berry curvature gives opposite Chern numbers for the two Weyl pairs, with \(C_{s}=-1\) along X--M and \(C_{s}=+1\) along Y--M, leading to a quantized spin Hall conductivity plateau inside the SOC-induced gap. The edge spectrum further confirms this topological character by showing counterpropagating edge states, demonstrating the coexistence of \(d\)-wave altermagnetism and topological states in Fe$_2$Se$_2$H monolayer.

\vspace{0.1cm}
\paragraph*{Functionalization-controlled transport.}
Having established the functionalization-induced altermagnetic spin splitting in \(\mathrm{Fe_2Se_2H}\), we next examine whether this symmetry-driven electronic structure can be translated into a measurable tunneling response. As a preliminary indication, we calculate the spin-resolved transmission along the [100] transport direction for pristine FeSe and functionalized Fe$_2$Se$_2$H, as shown in Fig.~S4. In pristine FeSe, the spin-up and spin-down transmission curves nearly overlap, consistent with its spin-degenerate conventional AFM band structure. By contrast, Fe$_2$Se$_2$H exhibits a pronounced spin-dependent transmission near the Fermi level, indicating that surface functionalization introduces an efficient spin-selective tunneling channel.

\begin{figure}[t!]
	\centering
	\includegraphics[width=0.48\textwidth]{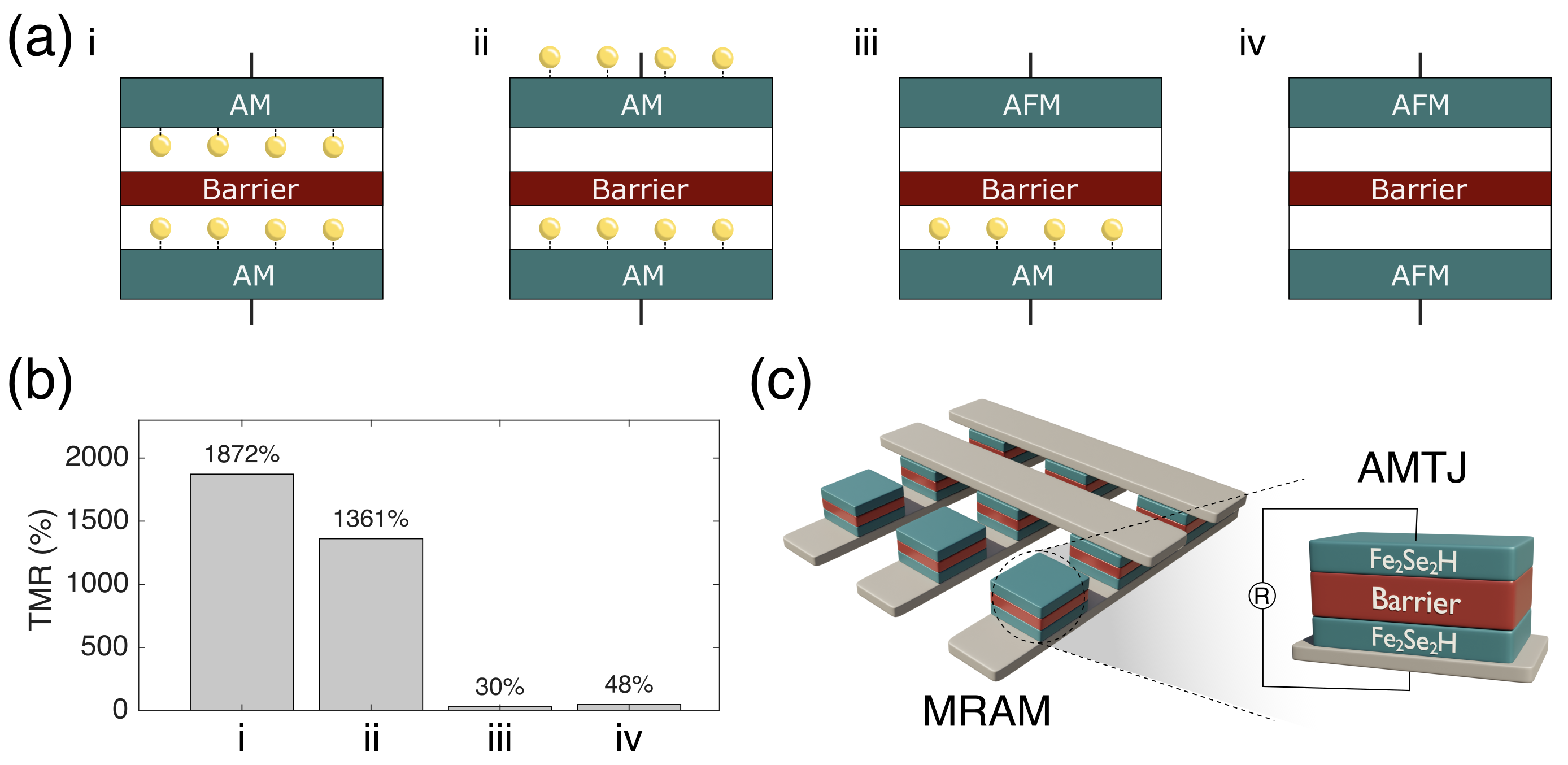}
	\caption{Surface functionalization-controlled tuning of TMR in AMTJ. (a) Schematic configurations of four-type hydrogenation at the barrier interfaces under PC and APC configurations. (b) Corresponding TMR ratios for different hydrogenation configurations. (c) Schematic illustration of AMTJs integrated into magnetoresistive random-access memory (MRAM) architectures. The device consists of Fe$_2$Se$_2$H electrodes separated by a GeF$_4$ tunneling barrier. 
	}
	\label{Figure4}
\end{figure}

To evaluate the device-level consequence of the surface functionalization-induced altermagnetism, we construct a vertical \(\mathrm{Au/Fe_2Se_2H/GeF_4/Fe_2Se_2H/Au}\) tunnel junction, as shown in Fig.~\ref{Figure3}(a). In this junction, the two Fe$_2$Se$_2$H monolayers act as spin pinned and free layers, while the \(\mathrm{GeF_4}\) layer serves as the nonmagnetic tunneling barrier and the Au layers function as metallic electrodes. The transport response is examined by comparing two relative N\'eel vector configurations of the \(\mathrm{Fe_2Se_2H}\) spin filters: the parallel configuration (PC), in which the two N\'eel vectors are aligned, and the antiparallel configuration (APC), in which they are reversed with respect to each other. The calculated energy-dependent transmission curves are shown in Fig.~\ref{Figure3}(b). Around the Fermi level, the transmission in the PC is much larger than that in the APC, giving rise to a sizable TMR of \(1.87\times10^3\%\). This value is significantly higher than those reported for many AM-based devices~\cite{Shao2021,10.1063/5.0278985,PhysRevB.108.024410}, demonstrating the strong potential of functionalized \(\mathrm{Fe_2Se_2H}\) for high-performance spintronic applications.

The microscopic origin of this large TMR is further revealed by the \(k_\parallel\)-resolved transmission maps in Fig.~\ref{Figure3}(c). In the PC, the high-transmission regions are strongly anisotropic and spin selective: the spin-up and spin-down channels appear in different sectors of the 2D Brillouin zone, forming complementary transmission windows. Because the two \(\mathrm{Fe_2Se_2H}\) layers share the same N\'eel-vector orientation, these spin-resolved windows are well registered across the junction, and their contributions add up to a large total transmission; While in the APC, the reversal of one N\'eel vector shifts the spin-selective tunneling windows out of registry, so the bright transmission lobes observed in the PC are largely suppressed and only weak residual channels remain. The resulting contrast between the open momentum-space channels in the PC and the blocked channels in the APC accounts for the high- and low-conductance states, respectively, and identifies momentum-selective altermagnetic spin filtering as the microscopic origin of the large TMR in the \(\mathrm{Fe_2Se_2H}\)-based AMTJ.
 
Surface-functionalization engineering offers an additional degree of freedom for designing AMTJ, because different adsorption geometries can directly change the layer composition, interfacial symmetry, and spin-filtering efficiency of the device. To clarify how functionalization affects tunneling transport, we compare four representative junction configurations, Fig.~\ref{Figure4}(a). Configurations i and ii contain two functionalized altermagnetic Fe$_2$Se$_2$H layers, but differ in the relative positions of the surface H atoms with respect to the tunneling barrier. In contrast, configurations iii and iv involve partially or fully unfunctionalized AFM FeSe layers, thereby weakening or removing the altermagnetic spin-selective transport channel. The calculated TMR ratios are summarized in Fig.~\ref{Figure4}(b). The fully functionalized AMTJ in configuration I exhibits the largest TMR of \(1872\%\), while configuration ii still maintains a large TMR of \(1361\%\). In sharp contrast, configurations iii and iv show much smaller TMR values of only \(30\%\) and \(48\%\), respectively. This comparison demonstrates that the large TMR originates from the functionalization-induced altermagnetic spin splitting rather than from conventional AFM tunneling. It also shows that not only the presence of functionalization, but also the adsorption geometry relative to the barrier, plays a decisive role in determining the transport efficiency.

These results establish surface functionalization as an effective control knob for altermagnetic tunnel transport. By selecting whether the FeSe layers are functionalized and how the functional groups are arranged at the interface, one can simultaneously tune the symmetry, spin splitting, and momentum-dependent spin filtering of the junction. As illustrated in Fig.~\ref{Figure4}(c), such functionalized \(\mathrm{Fe_2Se_2H}\)-based AMTJs can be naturally integrated into a magnetic random-access memory (MRAM) architecture, suggesting a feasible route toward device implementation. More broadly, the ability to encode tunnel-transport functionality through interfacial chemical design offers a practical guideline for developing scalable two-dimensional altermagnetic junctions.

\vspace{0.1cm}
In summary, we have proposed a symmetry-guided surface-functionalization strategy to transform 2D AFM into AM. By reshaping the local bonding environment, surface functionalization breaks the $I$ and $M_z$ symmetries while preserving the $R$-related symmetry connecting opposite-spin sublattices, thereby enabling a AFM-to-AM transition. Taking monolayer FeSe as a representative platform, we show that single-sided hydrogenation, oxidation, and fluorination can induce sizable \(d\)-wave altermagnetic spin splitting, with \(\mathrm{Fe_2Se_2H}\) further exhibiting nontrivial topological states. At the device level, an \(\mathrm{Fe_2Se_2H}\)-based AMTJ displays a large TMR of \(1.87\times10^3\%\), originating from momentum-selective altermagnetic spin filtering, while different functionalization configurations demonstrate that adsorption geometry provides an effective control knob for tunnel transport. By exploiting the chemical flexibility of surface functionalization, this strategy enables controllable design of altermagnetism in atomically thin materials through the choice of functional groups, adsorption geometries, and interfacial configurations. It therefore points to functionalized altermagnets as versatile building blocks for high-TMR spintronic devices, magnetic random-access memory, spin filtering, and topological spin transport.

\paragraph*{Calculation methods.}
\vspace{0.1cm}
The first-principles calculations based on density functional theory (DFT) are performed using the projector augmented-wave (PAW) pseudopotential method as implemented in the VASP code~\cite{PhysRevB.50.17953}. The Perdew-Burke-Ernzerhof generalized gradient approximation (GGA-PBE) is employed to describe the exchange and correlation functional. The plane-wave energy cutoff is set to $500$ eV, with a total energy convergence criterion of $1\times10^{-6}$ eV. All atoms in the unit cell are allowed to relax until the Hellmann-Feynman force on each atom was less than $0.01$ eV/\AA{}. In the case of two-dimensional films, a vacuum space exceeding $20$ \AA{} is implemented to eliminate interactions between adjacent slabs. The GGA+$U$ method with an effective Hubbard $U$ value of $0.8$ eV is used to address the correction effects of $d$ electrons of Fe atoms~\cite{yuan2018edge,luo2022fragile}. Phonopy code is adopted to calculate the phonon dispersion curves. The DFT-D3 interaction is taken into account. 

The spin transport properties are calculated in QuantumATK package, based on DFT and the nonequilibrium Green's function (NEGF)~\cite{PhysRevB.65.165401}. The norm-conserving PseudoDojo pseudopotentials within a linear combination of atomic orbitals (LCAO) framework is utilized to address electron exchange and correlation interactions. The wave functions are expanded using a double-zeta basis set, with a cutoff energy of $120$ Hartree for the calculations of the constructed devices. $7\times7\times210$ and $121\times121$ $\vb{k}$-point meshes are employed to the self-consistent and spin-dependent transmission calculations, which are dense sufficient to get the accurate results. The spin-dependent conductance of the tunnel junction per unit cell area is calculated as follows~\cite{landauer1970electrical}:
\begin{equation}
G^\sigma = \frac{e^2}{h} \frac{1}{N} \sum_{\vb{k}} T_{\vb{k}}^\sigma(E)
\end{equation}
where $T^\sigma(E)$ is the $\vb{k}$-point averaged transmission function at energy $E$ and $T_{\vb{k}}^\sigma(E)$ is the $\vb{k}$-resolved transmission function with $\sigma = (\uparrow, \downarrow$).  The TMR ratio is calculated by the formula~\cite{yuasa2004giant}:
\begin{equation}
\text{TMR} = \frac{G_{\text{PC}} - G_{\text{APC}}}{G_{\text{APC}}} \times 100\%
\end{equation}
where $G_{\text{PC}}$ and $G_{\text{APC}}$ are the total spin-dependent conductance of parallel configuration (PC) and antiparallel configuration (APC) states, which are defined as:
\begin{equation}
G_{\text{PC}} = G_{\text{PC}}^\uparrow + G_{\text{PC}}^\downarrow
\end{equation}
\begin{equation}
G_{\text{APC}} = G_{\text{APC}}^\uparrow + G_{\text{APC}}^\downarrow
\end{equation}
where $T_{\text{PC}}^\uparrow$ and $T_{\text{PC}}^\downarrow$ represent the spin-up and spin-down transmission coefficients at the Fermi level when the MTJs are in PC state, and $T_{\text{APC}}^\uparrow$ and $T_{\text{APC}}^\downarrow$ represent the spin-up and spin-down transmission coefficients at the Fermi level when the MTJs are in APC state.


\vspace{1em}
\indent{$Acknowledgements$}---This work is supported by the Zhejiang Provincial Natural Science Foundation of China (LR25A040001), the Zhejiang Provincial Leading Innovative and Entrepreneurial Team Project (2025R01017), the National Natural Science Foundation of China (12474155, 12504108), the Zhejiang Leading Goose (Lingyan) Project (2026C02A2013(SD2)), and the China Postdoctoral Science Foundation (2025M773440). The computational resources for this research were provided by the High Performance Computing Platform at the Eastern Institute of Technology, Ningbo. 

\vspace{2mm}
Z. Cui, Z. Y. Zhu and  B. W. Hao contributed equally to this work.

\bibliography{Reference}

\end{document}